\documentstyle[12pt, epsf]{article}

\begin{document}
\newcommand{\be}{\begin{equation}}
\newcommand{\ee}{\end{equation}}
\newcommand{\lll}{\lambda}
\def\Journal#1#2#3#4{{#1} {\bf #2}, #3 (#4)}

\newcommand{\A}{\alpha}
\newcommand{\B}{\beta}
\newcommand{\T}{\theta}
\newcommand{\Ep}{\epsilon}
\newcommand{\beq}{\begin{equation}}
\newcommand{\eeq}{\end{equation}}
\newcommand{\fr}{\frac}
\newcommand{\beqn}{\begin{eqnarray}}
\newcommand{\eeqn}{\end{eqnarray}}
\newcommand{\G}{\gamma}
\newcommand{\D}{\delta}
\renewcommand{\P}{\phi}
\newcommand{\intl}{\int\limits_{1}^{\infty}}
\renewcommand {\L}{\lambda}
\newcommand{\pt}{\partial}

\newcommand{\bq}{\bar q_A}
\newcommand{\tq}{\tilde q_A}
\newcommand{\btq}{\bar{\tilde q}^A}
\newcommand{\fa}{\varphi^A}
\newcommand{\bfa}{\bar \varphi_A}

\newcommand{\none}{{\cal N}=1}                            
\newcommand{\ntwo}{{\cal N}=2}                            

\def\st{\scriptstyle}
\def\sst{\scriptscriptstyle}
\def\mco{\multicolumn}
\def\epp{\epsilon^{\prime}}
\def\vep{\varepsilon}
\def\ra{\rightarrow}
\def\al{\alpha}
\def\ab{\bar{\alpha}}
\def\bea{\begin{eqnarray}}
\def\eea{\end{eqnarray}}

\begin{flushright}
\begin{tabular}{l}
ITEP-TH-28/07\\
\end{tabular}
\end{flushright}

\vskip1cm

\centerline{\large
\bf Magnetic strings in Lattice QCD as  Nonabelian Vortices }

\vspace{1.5cm}

\centerline{\sc A.Gorsky$^{a}$ and V.Zakharov$^{a,b}$ }

\vspace{0.5cm}

\centerline{a.\ \it Institute of Theoretical and Experimental Physics }
\centerline{\it  B. Cheremushkinskaya ul. 25, 117259 Moscow, Russia}
\vspace{0.5cm}
\centerline{b.\ \it NFN - Sezione di Pisa,
Largo Pontecorvo, 3, 56127 Pisa, Italy}

\def\thefootnote{\fnsymbol{footnote}}%
\vspace{4.5cm}

\centerline{\large \bf {Abstract}}
\vspace{1.0cm}
Lattice studies indicate  existence
of  magnetic strings
in  QCD vacuum. We argue that
recently found nonabelian strings with rich
worldsheet dynamics provide a proper pattern
for the strings observed on the lattice. In particular,
within this pattern we explain
the localization of the monopole-antimonopole pairs on the
magnetic string worldsheet and  the negative contribution of the magnetic
strings into the
vacuum energy and gluon condensate. We suggest the D2
brane realization of the magnetic string which explains the temperature
dependence of its tension.

\newpage
\section{Introduction}

Explanation of the QCD vacuum structure  remains
a challenging problem.
Recently some progress has been
made in the lattice studies and their interpretation \cite{lattice}. In
particular
the essential contribution from the 2d surfaces (strings)
and 3d volumes(domain walls) with some unusual properties
to the vacuum characteristics has been found.
For our purposes,  the
key properties of the magnetic strings observed on the
lattice  can be summarized as follows (for references see
Section 4 below and reviews \cite{lattice}:
\begin{itemize}

\item{
The tension of the magnetic string vanishes below the critical temperature
and they percolate through the vacuum, forming a kind of a vacuum condensate }

\item{
The worldsheet of magnetic string is populated
by monopole-antimonopole pairs }

\item{
Above the temperature of the deconfinement phase transition
magnetic string becomes tensionful }
 \end{itemize}

Most recently, it was argued that
\begin{itemize}
\item{ The magnetic strings become
a component of Yang-Mills plasma \cite{chernodub}}
\end{itemize}
and the first measurements
indicate surprisingly  that
 \begin{itemize}
\item{  The contribution of the  strings to
the gluon condensate and 4d bulk
vacuum energy is opposite in sign compared to its total value \cite{naka}}
\end{itemize}

A natural question concerns the very existence of
strings with such properties in the continuum theory.
The goal of this note is to argue that nonabelian magnetic
strings found recently in the SUSY gauge theories naturally provide
the desired pattern.  We are not aiming to prove rigorously
that  nonabelian strings populate QCD vacuum. However
our considerations
clearly indicate that this kind of object fits perfectly
the lattice data.

The nonabelian strings which are essentially
twisted $Z_N$ strings with orientational moduli have been first found in
SUSY context \cite{hanany,auzzi}.  However, later it was recognized
that they do exist in non SUSY theories as well \cite{gsy} (see
\cite{tong,sy2007,sakai} for reviews).
The key property of the nonabelian strings which distinguishes
them from the other objects discussed in this
context is
highly nontrivial worldsheet theory which in the
simplest examples can be identified with $CP(N -1)$
sigma model. Moreover it was found that kinks on
the worldsheet are nothing else but the 4d monopoles "`trapped"'
by the string \cite{tonmon,sy2004}. In nonsupersymmetric case
$CP(N-1)$ worldsheet
theory is in the confinement phase \cite{witten} so that  only
kink-antikink pairs exist which parallels the lattice
QCD observations. It was also argued recently that nonabelian
strings could play an essential role in the Seiberg duality
\cite{sy07,eto}.

As is mentioned above, the very recent lattice data
indicates that magnetic strings
contribute to the vacuum energy and gluon condensate
with the unexpected sign \cite{naka}.
On the other hand, it was found long time ago
\cite{rev} that vacuum energies in 4d gauge theories
and 2d  $CP(N -1)$ sigma model have opposite signs.
We argue  that this old observation  provides a pattern for
an interpretation of the lattice data \cite{naka}.

The lattice data suggests that the tension of the magnetic string
is zero below the deconfinement temperature $T_c$ and the question is
whether the nonabelian strings share this property.
To get insight into the problem
we will use the brane realization of the nonabelian string
as D2 brane in the particular supergravity background. Within this picture we
argue that
interpretation of the magnetic string as the wrapped
D2 brane explains the temperature dependence of the
tension. The crucial point is that the worldsheet
action on the magnetic string consists of two parts:
"`space"' part presumably involving the Nambu-Goto
type contribution as well as "`internal"' $CP(N-1)$
part responsible for the rich structure
of the worldsheet theory. The vanishing of the "`space"' tension
of the string below the critical temperature
does not imply the vanishing of "`internal"' part which is still
responsible for   the nontrivial dynamics at small temperature.
The change of the background above the critical temperature
results in an  interesting phenomenon that
the properties of the time- and space-oriented magnetic strings
become different.

The paper is organized as follows. In Section 2
we explain the construction of the nonabelian string solution in the simple model.
In Section 3 we argue that
the nonabelian string pattern
explains the negative contributions to the vacuum energy and the
gluon condensate and temperature dependence of the magnetic string tension.
In Section 4 we briefly compare our picture with the
available lattice data on the magnetic strings.
Section 5 involves the brief discussion on the results obtained and some
unsolved issues.

\section{Nonabelian strings}
Here we review the simplest model which can be used
to analyze nonabelian strings.   The gauge group of the model
is SU($N)\times$U(1). Besides SU($N$) and U(1)
gauge bosons
the model contains $N$ scalar fields charged with respect to
$U(1)$ which form $N$ fundamental representations of SU($N$).
It is convenient to write these fields as
$N\times N$ matrix $\Phi =\{\varphi^{kA}\}$
where $k$ is the SU($N$) gauge index while $A$ is the flavor
index,
\beq
\Phi =\left(
\begin{array}{cccc}
\varphi^{11} & \varphi^{12}& ... & \varphi^{1N}\\[2mm]
\varphi^{21} & \varphi^{22}& ... & \varphi^{2N}\\[2mm]
....&...&...&...\\[2mm]
\varphi^{N1} & \varphi^{N2}& ... & \varphi^{NN}
\end{array}
\right)\,.
\label{phima}
\eeq
The action of the model has the form\,
\beqn
S &=& \int {\rm d}^4x\left\{\frac1{4g_2^2}
\left(F^{a}_{\mu\nu}\right)^{2}
+ \frac1{4g_1^2}\left(F_{\mu\nu}\right)^{2}
 \right.
 \nonumber\\[3mm]
&+&
 {\rm Tr}\, (\nabla_\mu \Phi)^\dagger \,(\nabla^\mu \Phi )
+\frac{g^2_2}{2}\left[{\rm Tr}\,
\left(\Phi^\dagger T^a \Phi\right)\right]^2
 +
 \frac{g^2_1}{8}\left[ {\rm Tr}\,
\left( \Phi^\dagger \Phi \right)- N\xi \right]^2
 \nonumber\\[3mm]
 &+&\left.
 \frac{i\,\theta}{32\,\pi^2} \, F_{\mu\nu}^a \tilde{F}^{a\,\mu\nu}
 \right\}\,,
\label{redqed}
\eeqn
where $T^a$ stands for the generator of the gauge SU($N$),
\beq
\nabla_\mu \, \Phi \equiv  \left( \partial_\mu -\frac{i}{\sqrt{ 2N}}\; A_{\mu}
-i A^{a}_{\mu}\, T^a\right)\Phi\, ,
\label{dcde}
\eeq
and $\theta$ is the vacuum angle. The action (\ref{redqed})
represents a truncated bosonic sector of the \ntwo \ SUSY model. The last
term in the second line
forces $\Phi$ to develop a vacuum expectation value (VEV) while the
last but one term
forces the VEV to be diagonal,
\beq
\Phi_{\rm vac} = \sqrt\xi\,{\rm diag}\, \{1,1,...,1\}\,.
\label{diagphi}
\eeq

We assume that the parameter $\xi$ to be large,
\beq
\sqrt{\xi}\gg \Lambda_4,
\label{weakcoupling}
\eeq
where $\Lambda_4$ is the scale of the four-dimensional theory (\ref{redqed}).
That is we are in the weak coupling regime as both couplings $g^2_1$ and
$g^2_2$
are frozen at a large scale.

The  vacuum field (\ref{diagphi}) results in  the spontaneous
breaking of both gauge and flavor SU($N$)'s.
A diagonal global SU($N$) survives
\beq
{\rm U}(N)_{\rm gauge}\times {\rm SU}(N)_{\rm flavor}
\to {\rm SU}(N)_{\rm diag}\,.
\eeq
yielding color-flavor locking  in the vacuum.

To describe the topological argument providing the stability
of the string one can
combine the $Z_N$ center of SU($N$) with the elements $\exp (2\pi i
k/N)\in$U(1)
to get a topologically stable string solution
possessing both windings, in SU($N$) and U(1). In other words,
\beq
\pi_1 \left({\rm SU}(N)\times {\rm U}(1)/ Z_N
\right)\neq 0\,.
\eeq
and this nontrivial topology amounts to winding
of just one element of $\Phi_{\rm vac}$,  for instance,
\beq
\Phi_{\rm string} = \sqrt{\xi}\,{\rm diag} ( 1,1, ... ,e^{i\alpha (x) })\,,
\quad x\to\infty \,.
\label{ansa}
\eeq
These strings can be called elementary $Z_N$ strings;
their tension is $1/N$-th of that of the ANO string.
The ANO string can be viewed as a bound state of
$N$ $Z_N$ strings.

The $Z_N$ string solution
can be written as
follows \cite{auzzi}:
\beqn
\Phi &=&
\left(
\begin{array}{cccc}
\phi(r) & 0& ... & 0\\[2mm]
....&...&...&...\\[2mm]
0& ... & \phi(r)&  0\\[2mm]
0 & 0& ... & e^{i\alpha}\phi_{N}(r)
\end{array}
\right) ,
\nonumber\\[5mm]
A^{{\rm SU}(N)}_i &=&
\frac1N\left(
\begin{array}{cccc}
1 & ... & 0 & 0\\[2mm]
....&...&...&...\\[2mm]
0&  ... & 1 & 0\\[2mm]
0 & 0& ... & -(N-1)
\end{array}
\right)\, \left( \pt_i \alpha \right) \left[ -1+f_{NA}(r)\right] ,
\nonumber\\[5mm]
A^{{\rm U}(1)}_i &=& \frac{1}{N}\,
\left( \pt_i \alpha \right)\left[1-f(r)\right] ,\qquad A^{{\rm U}(1)}_0=
A^{{\rm SU}(N)}_0 =0\,,
\label{znstr}
\eeqn
where $i=1,2$ labels coordinates in the plane orthogonal to the string
axis and $r$ and $\alpha$ are the polar coordinates in this plane. The profile
functions $\phi(r)$ and  $\phi_N(r)$ determine the profiles of the scalar
fields,
while $f_{NA}(r)$ and $f(r)$ determine the SU($N$) and U(1) fields of the
string solutions, respectively. These functions satisfy the following
boundary conditions:
\beqn
&& \phi_{N}(0)=0,
\nonumber\\[2mm]
&& f_{NA}(0)=1,\;\;\;f(0)=1\,,
\label{bc0}
\eeqn
at $r=0$, and
\beqn
&& \phi_{N}(\infty)=\sqrt{\xi},\;\;\;\phi(\infty)=\sqrt{\xi}\,,
\nonumber\\[2mm]
&& f_{NA}(\infty)=0,\;\;\;\; \; f(\infty) = 0
\label{bcinfty}
\eeqn
at $r=\infty$.
These profile functions satisfy the first-order differential equations,
namely,
\beqn
&&
r\frac{d}{{d}r}\,\phi_N (r)- \frac1N\left( f(r)
+  (N-1)f_{NA}(r) \right)\phi_N (r) = 0\, ,
\nonumber\\[4mm]
&&
r\frac{d}{{ d}r}\,\phi (r)- \frac1N\left(f(r)
-  f_{NA}(r)\right)\phi (r) = 0\, ,
\nonumber\\[4mm]
&&
-\frac1r\,\frac{ d}{{ d}r} f(r)+\frac{g^2_1 N}{4}\,
\left[(N-1)\phi(r)^2 +\phi_N(r)^2-N\xi\right] = 0\, ,
\nonumber\\[4mm]
&&
-\frac1r\,\frac{d}{{ d}r} f_{NA}(r)+\frac{g^2_2}{2}\,
\left[\phi_N(r)^2 -\phi(r)^2\right]  = 0\, .
\label{foe}
\eeqn

The tension of this elementary string is
\beq
T_1=2\pi\,\xi\, .
\label{ten}
\eeq
while the tension of
the ANO string is
\beq
T_{\rm ANO}=2\pi\,N\,\xi
\label{tenANO}
\eeq
which confirms its composite nature.

The elementary strings are essentially non-Abelian
since besides trivial translational
moduli, they give rise to moduli corresponding to spontaneous
breaking of a non-Abelian symmetry. Indeed, while the ``flat"
vacuum is SU($N$)$_{\rm diag}$ symmetric, the solution (\ref{znstr})
breaks this symmetry
down.
This means that the worldsheet  theory of
the elementary string moduli
is the $CP(N-1)$ sigma model.

To obtain the non-Abelian string solution from the $Z_N$ string
(\ref{znstr}) we apply the diagonal color-flavor rotation  preserving
the vacuum (\ref{diagphi}).
It is useful to pass to the singular gauge where the scalar fields have
no winding at infinity, while the string flux comes from the vicinity of
the origin. In singular gauge we have
\beqn
\Phi &=&
U\left(
\begin{array}{cccc}
\phi(r) & 0& ... & 0\\[2mm]
....&...&...&...\\[2mm]
0& ... & \phi(r)&  0\\[2mm]
0 & 0& ... & \phi_{N}(r)
\end{array}
\right)U^{-1}\, ,
\nonumber\\[5mm]
A^{{\rm SU}(N)}_i &=&
\frac{1}{N} \,U\left(
\begin{array}{cccc}
1 & ... & 0 & 0\\[2mm]
....&...&...&...\\[2mm]
0&  ... & 1 & 0\\[2mm]
0 & 0& ... & -(N-1)
\end{array}
\right)U^{-1}\, \left( \pt_i \alpha\right)  f_{NA}(r)\, ,
\nonumber\\[5mm]
A^{{\rm U}(1)}_i &=& -\frac{1}{N}\,
\left( \pt_i \alpha\right)   f(r)\, , \qquad A^{{\rm U}(1)}_0=
A^{{\rm SU}(N)}_0=0\,,
\label{nastr}
\eeqn
where $U$ is a matrix $\in {\rm SU}(N)$. This matrix parameterizes
orientational zero modes of the string associated with flux embedding
into  SU($N$). The orientational
moduli encoded in the matrix $U$  were first
observed in \cite{hanany, auzzi}.

Let us discuss the worldsheet description of the nonabelian string.
It is important that there are two independent contribution
from "`space"' and "`internal"' terms. The space-time
action does not reduce purely to the Nambu-Goto term
which is only the first approximation term. The corresponding
tension is proportional to $\xi$.
To obtain the   kinetic term in the "`internal"' action we follow the standard
logic in the derivation of the low-energy
action in the moduli approximation. That is we substitute our solution, which
depends
on the moduli $ n^l$, in the action , assuming  that
the fields acquire a dependence on the coordinates $x_k$ via $n^l(x_k)$.
Then we arrive at the $CP(N-1)$  sigma model (for details see  \cite{sy2007}),
\beq
S^{(1+1)}_{CP(N-1)}= 2 f\,   \int d t\, dz \,  \left\{(\pt_{k}\, n^*
\pt_{k}\, n) + (n^*\pt_{k}\, n)^2\right\}\,,
\label{cp}
\eeq
where the coupling constant $f$ is given by a normalizing integral
defined in terms of the string profile functions which yields

\beq
f= \frac{2\pi}{g_2^2}\,.
\label{betag}
\eeq
that is two-dimensional coupling constant is determined by the
four-dimensional non-Abelian coupling.

The relation between the four-dimensional and two-dimensional coupling
constants (\ref{betag}) is obtained  at the classical level. In quantum theory
both couplings run hence we have to specify a scale at which the relation
(\ref{betag}) takes place. The two-dimensional $CP(N-1)$ model
is
an effective low-energy theory good for the description of
internal string dynamics  at low energies,  much lower than the
inverse thickness of the string which, in turn, is given by $g_2\sqrt{\xi}$.
Therefore,
$g_2\sqrt{\xi}$ plays the role of a physical ultraviolet  cutoff in
(\ref{cp}). Below this scale, the
coupling $f$ runs according to its two-dimensional renormalization-group flow.

The sigma model (\ref{cp}) is asymptotically free hence at large
distances it gets into the strong coupling regime.  The  running
coupling constant  as a function of the energy scale $E$ at one loop is given
by
\beq
4\pi f = N\ln {\left(\frac{E}{\Lambda_{CP(N-1)}}\right)}
+\cdots,
\label{sigmacoup}
\eeq
where $\Lambda_{CP(N-1)}$ is the dynamical scale of the $CP(N-1)$
model. As was mentioned above,
the UV cut-off of the sigma model at hand
is determined by  $g_2\sqrt{\xi}$.
Hence,
\beq
\Lambda^N_{CP(N-1)} = g_2^N\, \xi^{N/2} \,\, e^{-\frac{8\pi^2}{g^2_2}} .
\label{lambdasig}
\eeq
In the bulk theory, due to the VEV's of
the scalar fields, the coupling constant is frozen at
$g_2\sqrt{\xi}$. There are no logarithms in the bulk theory
below this scale and the logarithms of the
world-sheet  theory take over.

The brane realization of the nonabelian strings can be captured
from the brane realization of the $CP(N-1)$ models. It corresponds
to the theory on the worldvolume of D2 brane in the background of two NS5
branes and $N$ D4 branes. The  brane geometry can be seen from
the D2 worldvolume perspective or $N$ D4 brane worldvolume perspective
providing the rationale for the relation between the physics of
2d $CP(N-1)$  model and   4d SQCD \cite{dorey}.

\section{Magnetic strings versus nonabelian strings}
\subsection{Monopole pairs on the worldsheet}

Let us show that the pattern of the nonabelian strings
provides the explanation of the properties of
the magnetic strings
observed on the lattice.  The first point we would like
to note is that from the discussion above it is clear that
monopole pairs
are present on the nonabelian string indeed.

The worldsheet theory is nonsupersymmetric $\sigma$-model which has
single vacuum state and the spectrum consists of kink-antikink
bound states \cite{witten}. These bound states can be identified
with monopole-antimonopole bound states from the four-dimensional
viewpoint. The IR  scale $\Lambda_{CP}$ is generated in the worldsheet
theory
and can be related to the scale in the bulk theory.
The masses of the bound states in the theory are of order $\Lambda_{CP}$
and they cannot be found exactly since worldsheet theory is in strong
coupling regime. In the SUSY setup one can consider massive
flavors yielding the quasiclassical picture of the bound states.
In the nonSUSY case we have no such simple argumentation.
Note however that the monopoles in the
Higgs phase
on the string worldsheet are smoothly related to the
T'Hooft-Polyakov monopoles via the continuous
deformation in the parameter space (see a recent
discussion in \cite{gsy2007}).

If one introduces $\theta$ term in the bulk theory then
due to Witten effect monopoles acquires the electric charge
and becomes a dyon. Similar picture happens on the worldsheet
as well. The $\theta$ term penetrates into the worldsheet theory
and kink in the worldsheet theory acquires the global charge.

\subsection{Vacuum energy and gluon condensate}
In view of the recent lattice measurements \cite{naka} of contribution
of magnetic strings into Yang-Mills plasma energy
we will consider the energy issue in the context of the nonabelian strings.
There are two contributions to the energy associated with
dynamics in space-time and internal space, respectively.
These contributions can be treated separately.
Our basic observation is that
the contribution from
the internal, $CP(N-1)$ part  is in fact negative and opposite in sign
to its total value.

First, note that vacuum energy (at vanishing temperature)
in the Yang-Mills theory  is related to
the conformal anomaly:
\beq
E_{vac}^{YM}=
\frac{1}{4}<0|\theta_{\mu\mu}^{YM}|0>=<0|-\frac{b_0\alpha_S}{32\pi} TrG^2|0>
\eeq
where $b_0$ is the beta function coefficient. Similar
relation holds in $CP(N-1)$ model as well. Namely
\beq
E_{vac}^{CP}= \frac{1}{2}<0|\theta_{\mu\mu}^{CP}|0>=
\frac{N}{8\pi}\Lambda_{CP}^2
\eeq
The gluon condensate $<TrG^2>$
gets contribution from the nonabelian strings
since the internal tension is proportional to inverse gauge
coupling
\beq
<TrG^2>_{tot} \propto\frac{d}{d(1/g^2)}log Z \propto
<TrG^2>_{YM} + C_{CP} <TrG^2>_{CP}
\eeq
The two-dimensional contribution from the
nonabelian strings  comes from the vacuum expectation
value of two-dimensional
conformal anomaly in $CP(N-1)$ model which
has the opposite sign \cite{rev} compared to the total value
\beq
<TrG^2>_{CP}\propto <0|\theta_{\mu\mu}^{CP}|0> =
\frac{N}{8\pi}\Lambda_{CP}^2~~.
\eeq
The value of the dimensionful constant $C_{CP}$
is determined by the density of the strings and we
can not estimate its value at a moment.

Let us emphasize that we considered only the nonperturbative
contributions to the vacuum energy  which is determined by the
nonperturbative contribution to the conformal anomaly. The
contribution to the vacuum energy from the space part
of the string action vanishes since it is proportional to the
string tension.

\subsection{Low-energy Theorems and Dilaton}

Condensation of the magnetic string, observed on the lattice,
requires the consideration of the back reaction of the
single nonabelian string on the bulk fields. Below we
use the low-energy theorems in $CP(N-1)$ model to argue that
scalar mode on the worldsheet
contributes negatively to the mass squared of the
corresponding mode in four dimensions contrary to the
pseudoscalar case.

In the bulk theory, for any operator $A$ there holds
the dilatation Ward identity:
\beq
i\int d^4x <0|\theta_{\mu\mu}^{YM}(x) A(0)|0>= -d_{A} <0|A|0> ~~.
\eeq
where $d_{A}$ is the canonical dimension of the operator A.
This equation follows from the very fact of
asymptotic freedom in the theory.
Similar arguments apply to
the worldsheet theory and the corresponding  dilatation Ward identity
reads \cite{rev}:
\beq
i\int d^2x <0|\theta_{\mu\mu}^{CP}(x), A(0)|0>=-d_{A} <0|A|0>
\eeq
where we consider the correlator of the $\theta_{\mu \mu}$
with the arbitrary operator in the $CP(N-1)$
sigma model.

Some of the operators $A$ are of the special interest. Consider first the
operator
$A=TrG^2$ so that the corresponding  low-energy theorem reads as:
\beqn
i\int d^4x <0|TrG^2(x), TrG^2(0))|0>=S_{YM}(0)\propto <0|TrG^2|0>
\eeqn
In YM theory the r.h.s. is positive that is if we consider
the particle saturating S(0) it  has the positive mass. This particle naturally
could be related to the dilaton $\phi$ because of the standard coupling of the
dilaton $e^{\phi}TrG^2$.  On the other hand it is clear from the arguments
above that
this correlator has contribution from the string of the form
\beq
i\int d^2x <0|\theta_{\mu\mu}^{CP}(x), \theta_{\mu\mu}^{CP}(0)|0>=S_{CP}(0)
\eeq
The low-energy theorem yields   $S_{CP}<0$ \cite{rev}
which corresponds to  tachyonic contribution to the
particle in the intermediate state. The total
mass of scalar is positive while the stringy contribution is negative.

Let us compare the bulk-worldsheet interplay of the dilaton dynamics with the
similar consideration concerning axion \cite{axion}. It was shown in
\cite{axion}
that two-dimensional axion due to the mixing with photon is responsible for
deconfinement on the worldsheet. The reason is that because of this mixing
worldsheet
photon becomes massive and linear confinement disappears. On the other hand the
nonabelian string does not cause strong modification of the bulk dynamics and
results only on the axion emission halo around the string.

One can consider the correlator of the topological charge densities
\beqn
\frac{d^{2}logZ}{d^2{\theta}} =\int d^4 x<0|TrG \tilde{G}(x),TrG
\tilde{G}(0)|0>=P_{YM}(0)
\eeqn
which can be saturated by the axion in the intermediate state (we have no quarks). Since the four dimensional $\theta$ term penetrates into
the worldsheet theory
the $P(0)$ has the two-dimensional contribution
\beqn
i\int d^2 x <0|F(x) F(0)|0>=P_{CP}(0)
\eeqn
where $F=\epsilon_{\mu\nu}\partial_{\mu}A_{\nu}$
and $A_{\mu}$ is the auxiliary abelian gauge
field in $CP(N-1)$ model which acquires the mass at the
quantum level.
The contribution of the $P_{CP}$  to the total P(0) depends on the density
of the nonabelian strings but the crucial point now is
the sign of this contribution. Namely, it is known
\cite{rev} that the value of   $P_{CP}(0)$ is
positive and has the same sign as the total correlator. That is, we have
opposite influence of the
nonabelian strings on the dynamics of the dilaton and axion. The string tends
to decrease the
mass of scalar and acts oppositely in the pseudoscalar  case.

\subsection{Magnetic string tension from brane perspective}

Let us discuss the brane realization  of the magnetic string. To begin
with,   consider the weak-coupling
nonabelian string.  In the $N=2$ SQCD case
nonabelian string is perfectly identified as D2 brane
stretched between two NS5 branes displaced at the large distance
$\xi$ is some direction. According to the standard logic the
tension of the nonabelian  string turns out to be proportional
to $\xi$ that is quasiclassical analysis is reasonable.

The geometry of the nonSUSY QCD is not  established
well enough. However, the natural starting point is the
geometry provided by the set of D4 branes wrapped around one compact
dimensions \cite{wittenterm}. We shall consider the pure
gauge sector and does not discuss chiral matter in this note.

We shall assume the large $N_C$ limit and consider the supergravity
approximation.
In this approximation the geometry looks as $M_{10}=R_{3,1}\times D \times S^4$ and
the corresponding
metric reads as
$$
ds^2=(\frac{u}{R_0})^{3/2}(-dt^2 + \delta_{ij}dx^i dx^j +f(u)dx_4^2)+
(\frac{u}{R_0})^{-3/2}(\frac{du^2}{f(u)} +u^2 d\Omega_{4}^2)
$$
\be
e^{\Phi}=(\frac{u}{R_0})^4, \qquad F_4=\frac{3N_c\epsilon_4}{4\pi},\qquad
f(u)=1- (\frac{u_{\Lambda}}{u})^3
\eeq
where $R_0=(\pi g_s N_c)^{1/3}$ and
$R=\frac{4\pi}{3}(\frac{R_0^3}{u_{\Lambda}})^{1/2}$.
The coupling constant of Yang-Mills theory is related to the radius
of the compact dimension $R$ as follows
$$
g_{YM}^2=\frac{8\pi ^2 g_s l_s}{R}
$$

At zero temperature theory is in the confinement phase
and in the ($u,x_4$) coordinates we have the geometry
of the cigar with the tip at $u=u_{\Lambda}$. The D4 branes
are located along our D=4 geometry and are extended
along $x_4$ coordinate.
Let us emphasize that for the magnetic string we
discuss  the target space looks as $M_{10}\times CP(N-1)$
and involves the "internal" part.

Let us turn to our proposal for the magnetic string
within the brane setup. Adopting the nonabelian string
as the correct pattern we suggest that magnetic string
at strong coupling regime is the probe D2 brane wrapped
around $S_1$ parameterized by $x_4$ and its tension
is therefore proportional to the effective radius $R(u)$.
Due to the cigar geometry  this wrapping
is topologically unstable and the D2 brane shrinks to the tip
where its tension vanishes. That is, one immediately
reproduces the tensionless property of the magnetic string at zero
temperature.

The next point concerns the worldsheet theory on D2 brane.
We would like to see qualitatively the degrees of freedom corresponding
to $CP(N-1)$ model as well as the correct $\theta$-term in the
worldsheet Lagrangian. The D2 brane supports U(1) gauge
field on the worldvolume while the open D2-D4 strings provide
the matter necessary for the gauge formulation of $CP(N-1)$ model.
To trace the $\theta$ term  let us consider
the CS term on the D2 worldsheet
\beq
L_{CS}=\int d^3 x\ C_1\wedge F
\eeq
where $C_1$ is the R-R one-form field. Taking into account that
$\theta = \int dx_4\ C_1$ we reproduce the standard $\theta$-term
in $CP(N-1)$ model $\theta\int d^2 x\ F$.

The consideration of the finite $N_C$ case is much more subtle.
Let us consider finite $N_C$ D4 brane wrapped around $x_4$
coordinate and a single D2 brane wrapped around $x_4$ as well.
What could be the mechanism for the magnetic string condensation
for finite $N_C$? The natural conjecture is that the phenomena of dissolving
of p-brane inside (p+2)-brane \cite{gava} takes place here. Indeed
in our D2-D4 system we have proper brane dimensions and the
tachyonic mode of D2-D4 open string could lead to the D2 brane
condensation providing the additional
magnetic field  in four-dimensional  theory via the CS term
on the D4 worldsheet
\beq
L_{CS}=\int d^5 x\ C_3\wedge F
\eeq
induced by D2 brane  RR field.

However it is not clear how this D2-D4 tachyonic mode
could disappear at the critical temperature and this point needs for
the careful consideration. The possible stabilization mechanisms
of p-brane inside (p+2)-brane which could be
relevant in our context were discussed in \cite{d1d3} and are based
on the account of RR fields in the bulk.

We are not able to describe the string condensation
explicitly however one general comment is in order.
It is known that the condensation of the electric charges results
in confinement of the dual  magnetic charges. The natural
question concerns the nature of the dual object in our
case and their possible "`confinement"'. The magnetic
string is assumed to be  the source of the RR 2-form field $C_{\mu\nu}$
with non-vanishing curvature three-form $B_3=dC_2$. The duality relation
\beq
*B=da
\eeq
imply that the dual field is pseudoscalar axion .
Hence the dual objects which interact with the dual field
are carries of the topological charge and we could expect
"`confinement of the carriers of the topological charge"' upon the
magnetic string condensation.

\subsection{Temperature dependence of the magnetic string}

The crucial test of the proposal concerns the temperature
dependence of the magnetic string. We have argued
above that at zero and small temperatures the
cigar geometry in ($x_4,u$) plane amounts to the vanishing
tension of the magnetic string since the radius of
the circle D2 brane wrapped around shrinks to zero.
However the magnetic string  becomes
tensionful above the critical temperature $T_c$ of the
deconfinement phase transition. How  the change of two
regimes happens?

The key point is that in the temperature
case there are two backgrounds with the asymptotic topology of
$R^3\times S^1_{\tau} \times S^1 \times S^4$, where
$\tau$ is the Wick rotated time coordinate $\tau=it$,
$\tau\propto \tau + \beta$.
One background corresponds to the  analytic continuation
of the previous one while the second corresponds
to the interchange of $\tau$ and $x_4$, that is
the warped factor is attached to the $\tau$ coordinate
and cigar geometry emerges in the ($\tau,u$) plane instead
of ($x_4,u$) plane which now exhibits the cylinder geometry, see Figure. It was
shown in
\cite{temperature} by calculation of the free energies that above $T_c$ the
second background dominates.

\begin{figure}
\epsfxsize=15cm
\centerline{\epsfbox{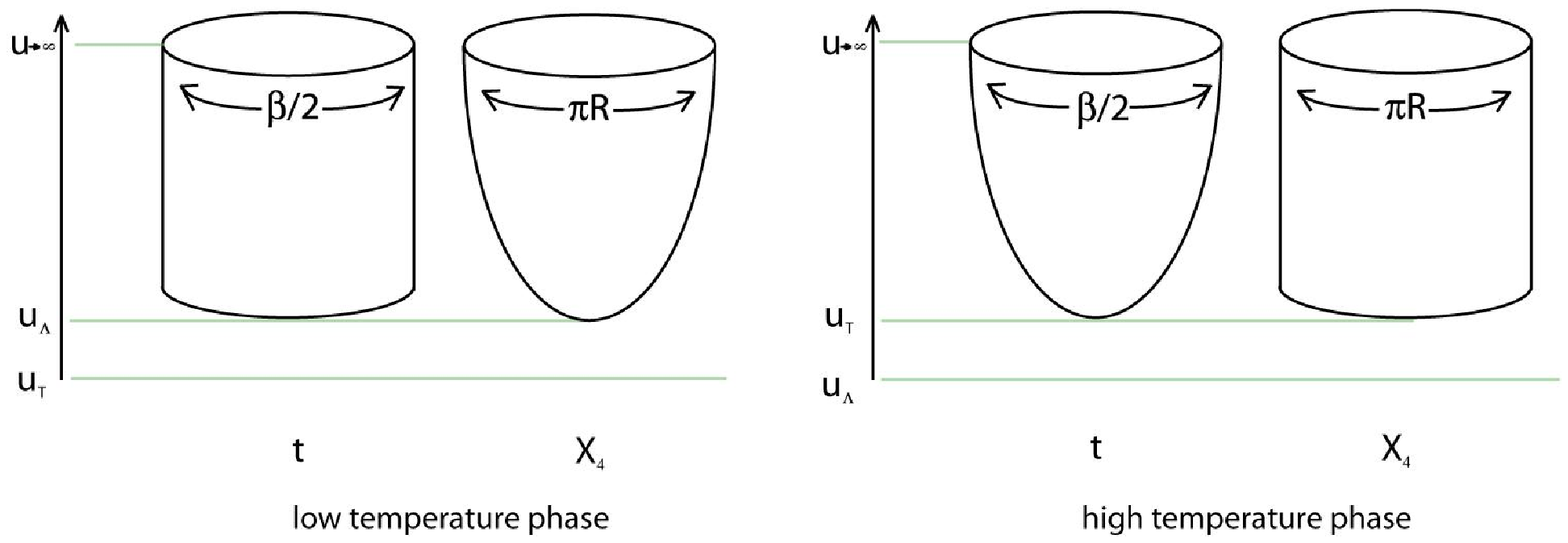}}
\end{figure}

Thus,  above $T=T_c$   one gets the
geometry of the cylinder in ($x_4,u$) and cigar in ($\tau,u$) so that the
 wrapping around $x_4$ is topologically
stable now and the magnetic string tension is
proportional the cylinder radius.
Moreover,
by  construction the D2 brane is wrapped around $x_4$ coordinate
but the rest two coordinates of the D2 brane can
fill the  different dimensions. If  both coordinates
of the magnetic string are transverse to the time direction
it does not feel the instability in the ($\tau,u$) cigar geometry
and behaves as S-string. On the other hand, if the magnetic
string is wrapped around $\tau$ coordinate it is unstable
in the cigar geometry and  shrinks along the $\tau$ coordinate
to zero. That is magnetic string extended in the time direction
looses one physical dimension above $T_C$ and, speaking somewhat loosely,
looks as a "`particle".
One could
say that vanishing tension below the critical temperature is "`traded for"' a
lost dimension above the critical temperature.

Let us emphasize that we have discussed above the
"`space"' tension corresponding to the linear density of
the energy which jumps at the phase transition point. On the
other hand the "`internal"' tension  defining the $CP(N-1)$
part of the action of the magnetic string $T_{int}=1/g^2$
goes smoothly at any temperature according to the
asymptotic freedom. Remark that the internal part of the action
follows from the open strings stretched between D2 brane
and the rest of D4 branes.

Note that the discussion in this section
is somewhat similar to the consideration in \cite{bergman}
of the role of the instantons in the similar geometry
which are represented by Euclidean D0 branes wrapped around
$x_4$. In that case it was argued that single instanton
is ill-defined below $T_c$ because of D0 brane instability
in the cigar geometry while above $T_c$ it is well defined
due to geometry of the cylinder.
The change of the
instanton role at the transition point corresponds
to the change from the Witten-Veneziano to t'Hooft
mechanisms of the solution to $U(1)$ problem.

\section{Lattice data}
\subsection{Lattice strings at zero temperature}

In this section we will provide a short guide to the literature
on lattice measurements relevant to the theoretical issues discussed
in this note.

Magnetic strings were introduced first in the context of the confinement
studies, as confining field configurations  and are known mostly as
`center vortices', for review and references see \cite{greensite}.
In particular it was found that the vortices percolate in the vacuum,
i.e. form an infinite cluster or a kind of condensate. Also, their total area
of the vortices is in physical
units,
\beq
(Area)_{total}~\sim~\Lambda_{{QCD}}^{2}V_{total}~~,
\eeq
where $V_{total}$ is the  total volume of the lattice.

For confinement, the transverse size of the strings is not crucial and
the strings were mostly thought of
as 'thick vortices'.
The fact that they are actually thin, two-dimensional surfaces was discovered
as a result of measurements of the distribution of nonabelian action
associated with the vortices \cite{gubarev}. The action turned to be singular
in the continuum limit,
\beq\label{singular}
(Action)_{{lattice}}~\sim~(Area)_{total}/a^{2}~~,
\eeq
where $a$ is the lattice spacing, $a\to 0$ in the continuum limit.  Moreover,
the
nonabelian field living on the surface is aligned, or 'trapped' to the surface.
It is these, thin strings which are relevant to our discussion.
Moreover, the strings are closed in the vacuum state but can be open
on an external 't Hooft line, for argumentation and references see
\cite{lattice}.
Hence, the name of `magnetic strings'.

Note the physical string tension is not directly related to the lattice
action but is to be rather calculated as a difference between
lattice action and entropy factors, see, e.g., \cite{ambjorn}.
It is difficult to check such a cancelation directly. The fact that
the physical tension for the lattice magnetic strings is vanishing in
the confining phase follows from the very existence of an infinite,
or percolating cluster of surfaces. Indeed, if the tension were not zero
only finite clusters could be observed, by virtue of the uncertainty principle.

Lattice monopoles are identified as closed trajectories, or particles,
for review see \cite{suzuki}.
Their lattice algorithmic  definition is independent of the definition
of the surfaces, or strings. Nevertheless, the lattice simulations reveal
that the monopole trajectories  lie in fact on the surfaces
\cite{giedt,gubarev}. The nonabelian fields associated with the monopoles
are also singular \cite{bornyakov} and are aligned with the surfaces
\cite{gubarev}.

All the lattice data on the magnetic strings are obtained with the standard
Wilson action and in fact refer mostly to pure Yang-Mills cases.
There is no direct explanation of the data within the Yang-Mills theory
itself. One can check, however, that the singular nonabelian fields
are just of the type which is in no contradiction
with the asymptotic freedom \cite{vz}.

\subsection{Lattice strings in the deconfinement phase}

We also considered strings at non-zero temperature and here we will provide
references to the lattice measurements at temperatures above the
deconfinement
phase transition.

The basic observation, made on the lattice \cite{langfeld,greensite}
is that at temperatures above the phase transition the strings become
time-oriented. The four-dimensional infinite percolating cluster is dissolved
and does not exist any longer. However, the percolation is not eliminated
altogether. Namely, in three-dimensional slices the four-dimensional strings
are
projected to lines. In case of the magnetic strings, the properties of these
lines depend crucially on whether one considers equal-time or
equal-space-coordinate slices. In case of equal-time slices the lines, which
are
intersections of the strings and of the 3d spaces, continue to percolate.
In case of the equal-space-coordinate slices, there is no
percolation at all.

Clearly, these lattice data are reproduced by the phenomenon
of a `missed dimension' discussed in detail in Section 3.5 in
the brane language.

\section{Discussion}

During the last thirty years the derivation of the
Mandestam's  qualitative
explanation of the confinement via the dual Meissner
effect was the main goal of the nonperturbative QCD studies.
The recent lattice data suggests that probably the picture is to
be modified and  condensation of the tensionless
magnetic strings takes place in QCD vacuum,
instead of the condensation of the
magnetic monopoles. If fact,
there is no deep contradiction between two scenarios.
Indeed the magnetic strings observed on the lattice
support the monopoles at their worldsheets. In other words,
condensation
of the strings implicitly assumes the condensation the
monopoles.
The monopoles become, however, particles living on a string, or in 2d instead
of ordinary particles living in 4d.

In this paper we conjecture that the strings observed
on the lattice follow the pattern of the nonabelian
strings with their rich worldsheet structure
supporting monopole-antimonopole pairs.
We have argued that this picture explains qualitatively
quite a few effects observed on the lattice in  pure Yang-Mills case.
Moreover, it turns out that the
interpretation of the magnetic strings as wrapped
D2 branes
fits perfectly with their properties.

What could we say about the wave function of the condensate?
Since the lattice data indicate  magnetic string  condensate
component in the vacuum  the
important point concerns the "`phase of the condensate wave function".
In other terms the question is what is  the remnant of the internal $CP(N-1)$-type
part of the action of the
individual string in the condensate. We could speculate that the
two-dimensional
$CP(N-1)$ model on the string worldsheet is promoted to
the four-dimensional $CP(N-1)$ sigma-model upon condensation.
This four-dimensional model supports the BPS stringy
solution which corresponds to the two-dimensional instanton
solution of $CP(N-1)$ model lifted to four-dimensions. This
presumably can be treated as the "`confinement of instantons"'
in the dual picture since individual instanton does not
exist from the topological reasons in 4d $CP(N-1)$ model.

In this paper we have focused on the pure Yang-Mills theory however
the generalization to QCD with fundamental quarks is possible. In
particular it is interesting to investigate the role of the
magnetic strings in the chiral properties of the theory. In the
brane setup we can add $N_f$ flavor branes and analyze
the dynamics in the Sakai-Sugimoto model \cite{ss} (see also \cite{before}
for the earlier papers). We have discussed magnetic strings that
is wrapped D2 branes only. However there are other
wrapped D4 and D6 branes in this setup which have an interesting
interpretation. These issues shall
be discussed elsewhere.

This project was started during the KITP at Santa Barbara
program ''QCD and  String Theory`` in the fall 2004 and we would like
to thank the organizers for the very creative atmosphere.
We would like to thank O. Andreev, M. Chernodub, A. DiGiacomo, F. Gubarev, M. Polikarpov,
M.Shifman, A. Vainshtein and A.Yung for the discussions and comments.
We are grateful to M. Chernodub and A. Nakamura for providing
us with the new lattice date prior the publication.
The work of A.G. was partly supported by grants RFBR-06-02-17382
and INTAS-05-1000008-7865.


\begin{thebibliography}{99}
\bibitem{lattice}
  V.~I.~Zakharov,   {\it `` Dual string from lattice Yang-Mills theory''},
 { AIP Conf. Proc.}  {\bf 756} (2005) 182,
 [ArXiv:hep-ph/050101];\\
 {\it  ``From confining fields on the lattice to higher dimensions in the continuum,''}
  {Brazilian J. Phys. } {\bf 37}, (2007) 165.
  [arXiv:hep-ph/0612342].



    \bibitem{chernodub}
  M.~N.~Chernodub and V.~I.~Zakharov,
  ``Magnetic component of Yang-Mills plasma,''
  Phys.\ Rev.\ Lett.\  {\bf 98}, 082002 (2007)
  [arXiv:hep-ph/0611228].\\
  M.~N.~Chernodub and V.~I.~Zakharov,
  ``Magnetic strings as part of Yang-Mills plasma,''
  arXiv:hep-ph/0702245.


\bibitem{naka}
 M.N. Chernodub, A. Nakamura,
  presented at  the workshop ``The Many Faces of Quantum Fields'',
Leiden, The Netherlands, April 2007, and to be published.

\bibitem{hanany}
  A.~Hanany and D.~Tong,
  ``Vortices, instantons and branes,''
  JHEP {\bf 0307}, 037 (2003)
  [arXiv:hep-th/0306150].

\bibitem{auzzi}
  R.~Auzzi, S.~Bolognesi, J.~Evslin, K.~Konishi and A.~Yung,
  ``Nonabelian superconductors: Vortices and confinement in N = 2 SQCD,''
  Nucl.\ Phys.\  B {\bf 673}, 187 (2003)
  [arXiv:hep-th/0307287].
\bibitem{gsy}
  A.~Gorsky, M.~Shifman and A.~Yung,
  ``Non-Abelian Meissner effect in Yang-Mills theories at weak coupling,''
  Phys.\ Rev.\  D {\bf 71}, 045010 (2005)
  [arXiv:hep-th/0412082].

\bibitem{tong}
  D.~Tong,
  ``TASI lectures on solitons,''
  arXiv:hep-th/0509216.

\bibitem{sy2007}
  M.~Shifman and A.~Yung,
  ``Supersymmetric Solitons and How They Help Us Understand Non-Abelian
Gauge Theories,''
  arXiv:hep-th/0703267.
\bibitem{sakai}
  M.~Eto, Y.~Isozumi, M.~Nitta, K.~Ohashi and N.~Sakai,
  ``Solitons in the Higgs phase: The moduli matrix approach,''
  J.\ Phys.\ A  {\bf 39}, R315 (2006)
  [arXiv:hep-th/0602170].


\bibitem{tonmon}
  D.~Tong,
  ``Monopoles in the Higgs phase,''
  Phys.\ Rev.\  D {\bf 69}, 065003 (2004)
  [arXiv:hep-th/0307302].

\bibitem{sy2004}
  M.~Shifman and A.~Yung,
  ``Non-Abelian string junctions as confined monopoles,''
  Phys.\ Rev.\  D {\bf 70}, 045004 (2004)
  [arXiv:hep-th/0403149].

\bibitem{rev}
  V.~A.~Novikov, M.~A.~Shifman, A.~I.~Vainshtein and V.~I.~Zakharov,
  ``Two-Dimensional Sigma Models: Modeling Nonperturbative Effects Of Quantum
  Chromodynamics,''
  Phys.\ Rept.\  {\bf 116}, 103 (1984)


\bibitem{witten}
  E.~Witten,
  ``Instantons, The Quark Model, And The 1/N Expansion,''
  Nucl.\ Phys.\  B {\bf 149}, 285 (1979).

\bibitem{sy07}
  M.~Shifman and A.~Yung,
  ``Confinement in N=1 SQCD: One Step Beyond Seiberg's Duality,''
  arXiv:0705.3811 [hep-th].
\bibitem{eto}
  M.~Eto, K.~Hashimoto and S.~Terashima,
  ``QCD String as Vortex String in Seiberg-Dual Theory,''
  arXiv:0706.2005 [hep-th].




\bibitem{axion}
  A.~Gorsky, M.~Shifman and A.~Yung,
  ``Nonabelian strings and axion,''
  Phys.\ Rev.\  D {\bf 73}, 125011 (2006)
  [arXiv:hep-th/0601131].
\bibitem{wittenterm}
  E.~Witten,
  ``Anti-de Sitter space, thermal phase transition, and confinement in  gauge
  theories,''
  Adv.\ Theor.\ Math.\ Phys.\  {\bf 2}, 505 (1998)
  [arXiv:hep-th/9803131].

\bibitem{dorey}
  N.~Dorey,
  ``The BPS spectra of two-dimensional supersymmetric gauge theories with
  twisted mass terms,''
  JHEP {\bf 9811}, 005 (1998)
  [arXiv:hep-th/9806056].\\
 N.~Dorey, T.~J.~Hollowood and D.~Tong,
  ``The BPS spectra of gauge theories in two and four dimensions,''
  JHEP {\bf 9905}, 006 (1999)
  [arXiv:hep-th/9902134].

\bibitem{temperature}
 O.~Aharony, J.~Sonnenschein and S.~Yankielowicz,
  ``A holographic model of deconfinement and chiral symmetry restoration,''
  Annals Phys.\  {\bf 322}, 1420 (2007)
  [arXiv:hep-th/0604161].

\bibitem{gsy2007}
  A.~Gorsky, M.~Shifman and A.~Yung,
  ``N = 1 supersymmetric quantum chromodynamics: How confined non-Abelian
  monopoles emerge from quark condensation,''
  Phys.\ Rev.\  D {\bf 75}, 065032 (2007)
  [arXiv:hep-th/0701040].



\bibitem{bergman}
  O.~Bergman and G.~Lifschytz,
  ``Holographic U(1)A and string creation,''
  JHEP {\bf 0704}, 043 (2007)
  [arXiv:hep-th/0612289].
\bibitem{gava}
 E.~Gava, K.~S.~Narain and M.~H.~Sarmadi,
  ``On the bound states of p- and (p+2)-branes,''
  Nucl.\ Phys.\  B {\bf 504}, 214 (1997)
  [arXiv:hep-th/9704006].
\bibitem{d1d3}
  E.~J.~Copeland, R.~C.~Myers and J.~Polchinski,
  ``Cosmic F- and D-strings,''
  JHEP {\bf 0406}, 013 (2004)
  [arXiv:hep-th/0312067].\\
   L.~Leblond and S.~H.~H.~Tye,
  ``Stability of D1-strings inside a D3-brane,''
  JHEP {\bf 0403}, 055 (2004)
  [arXiv:hep-th/0402072].
\bibitem{greensite}
J. Greensite, {\it``The Confinement problem in lattice gauge theory''},
{ Progr. Part. Nucl. Phys.} {\bf 51} (2003) 1, [arXiv:hep-lat/0301023].

\bibitem{gubarev}
F.~V.~Gubarev,A.~V.~Kovalenko, M.~I.~Polikarpov, S.~N.~Syritsyn,
V.~I.~Zakharov,
{\it `` Fine tuned vortices in lattice SU(2) gluodynamics''},
{Phys. Lett.} {\bf B574} (2003) 136, [arXiv:hep-lat/0212003].

\bibitem{ambjorn}
A.M. Polyakov, {\it ''Gauge Fields and Strings''},
Harvard Academic Publishers,  (1987);\\
J. Ambjorn, {\it ''Quantization of geometry''}, [arXiv:hep-th/9411179].

\bibitem{suzuki}
  M.N.Chernodub, F.V.Gubarev, M.I.Polikarpov, A.I.Veselov,
 {\it ``Monopoles in the Abelian Projection of Gluodynamics''},
 {Prog. Theor. Phys. Suppl.} {\bf  131} (1998) 309,
[arXiv:hep-lat/9802036];\\
 A. Di Giacomo,
 {\it `` Monopole condensation and color confinement
''}, {  Prog. Theor. Phys. Suppl.} {\bf 131} (1998) 161,
 [arXiv:hep-lat/9802008];\\
 T. Suzuki,
 {\it `` Low-energy effective theories from QCD''},
 { Prog. Theor. Phys. Suppl.} {\bf 131} (1998) 633.

\bibitem{giedt}
J. Ambjorn,  J. Giedt,  J. Greensite, S. Olejnik,
{\it`` Vortex structure versus monopole dominance in Abelian projected gauge
theory''},
{ JHEP} {\bf 0002} (2000) 033, [arXiv:hep-lat/9907021].

\bibitem{bornyakov}
V.G. Bornyakov, et al.,
 {\it``Anatomy of the lattice magnetic monopoles''},
{\it Phys. Lett.} {\bf B537} (2002) 291, [arXiv:hep-lat/0103032].

\bibitem{vz}
 V.I. Zakharov, {\it `` Nonperturbative match of ultraviolet renormalon''},
 [arXiv:hep-ph/0309178].

\bibitem{langfeld}
 M. Engelhardt, K. Langfeld, H. Reinhardt, O. Tennert,
  {\it ``Deconfinement in SU(2) Yang-Mills theory as a center vortex
percolation transition''},   Phys. Rev. {\bf D61} (2000)  054504,
[arXiv:hep-lat/9904004].

\bibitem{ss}
   T.~Sakai and S.~Sugimoto,
  ``Low energy hadron physics in holographic QCD,''
  Prog.\ Theor.\ Phys.\  {\bf 113}, 843 (2005)
  [arXiv:hep-th/0412141].\\
 T.~Sakai and S.~Sugimoto,
  ``More on a holographic dual of QCD,''
  Prog.\ Theor.\ Phys.\  {\bf 114}, 1083 (2006)
  [arXiv:hep-th/0507073].

\bibitem{before}
  A.~Karch and E.~Katz,
  ``Adding flavor to AdS/CFT,''
  JHEP {\bf 0206}, 043 (2002)
  [arXiv:hep-th/0205236].\\
 M.~Kruczenski, D.~Mateos, R.~C.~Myers and D.~J.~Winters,
  ``Towards a holographic dual of large-N(c) QCD,''
  JHEP {\bf 0405}, 041 (2004)
  [arXiv:hep-th/0311270].

\end{thebibliography}
\end{document}